\documentclass[12pt]{iopart}

\begin{document}
\title[Spin-hybrid-phonon resonance in anisotropic quantum dots]
{Spin-hybrid-phonon resonance in anisotropic quantum dots}

\author{V A Margulis and A V Shorokhov}

\address{Institute of Physics and Chemistry, Mordovian State
University, 430000 Saransk, Russia} \ead{shorokhovav@math.mrsu.ru}
\begin{abstract}
We have studied the absorption of electromagnetic radiation of an
anisotropic quantum dot taking into account the spin-flip
processes that is associated with the interaction of the electrons
with optical phonons. It is shown that these processes lead to the
resonance absorption. Explicit formula is derived for the
absorption coefficient. The positions of the resonances peaks are
found.
\end{abstract}


\submitto{\JPCM}

\maketitle

\section{Introduction}

Quantum dots support three main types of intraband resonances
associated with absorption of high frequency electromagnetic
radiation by electrons: (a) transitions between two energy states
caused by photon absorption; (b) transitions between two energy
states accompanied by simultaneous absorption and emission of a
phonon; (c) transitions between two energy states accompanied by
simultaneous scattering by impurities. The above-mentioned
processes were studied by the authors for the case of anisotropic
parabolic quantum dot \cite{Gey01,Mar02,Mar07}. The intraband
resonances are of particular interest for its possible
applications because quantitative understanding of optical
properties due to electron-photon, electron-phonon and
electron-impurity interactions is important for a successful
construction of optical devices based on quantum nanostructures
\cite{Ros02,Sha96}. In particular, the semiconductors quantum
lasers based on the quantum dots is the most perspective for using
as active medium for the new generation of injection lasers
\cite{Bim99} and for using in infrared detectors. The external
magnetic field lets us to control the working frequency of these
devices and the magnitude of absorption.

In the case (b), the absorption of a quantum $\hbar\omega$ of the
high-frequency field is accompanied by the absorption or emission
of an optical phonon. However the interaction of electrons with
phonons can lead to additional resonances die to spin-flip
processes \cite{Mat67}. As a rule this effect can arise in
semiconductors with a strong spin-orbital interaction, in
particular, in III-V compounds. These spin-flip processes can be
considered in the second-order perturbation theory in
electron-photon and electron-impurities perturbations using the
method developed in \cite{Bas65}.

In this paper we consider an anisotropic quantum dot located in a
$2D$ layer. We model the confining potential in the direction
perpendicular to the $2D$ layer using $\delta$-function potential.
The confining potential in the plane we model using parabolic
potential with characteristic frequencies $\Omega_x$ and
$\Omega_z$. The magnetic field $\bf{B}$ is directed along $y$-axes
to be perpendicular to the $2D$ layer.

Note that these resonance can be observed only if all levels
(including spin sublevels) are well-resolved and the photon
frequency is sufficiently monochromatic. Hence in what follows we
assume that the photon frequency is high ($\omega\tau\gg1$, $\tau$
is the relaxation time), the hybrid confinement is sufficiently
strong $\Omega_i\tau\gg1$, quantizing $\hbar\Omega_i\gg T$
($i=x,z$) and magnetic field is sufficiently strong
$\omega_c\tau\gg1$ ($\omega_c=eB/m^*c$ is the cyclotron frequency,
$m^*$ is the effective mass). In this case the transitions occur
between levels of the discrete spectrum.

Using the method suggested in Refs.\cite{Mat67,Bas65}, we find the
absorption coefficient by applying ordinary perturbation theory
for the interactions of electrons with the high-frequency field
$H_R$ and the lattice $H^{sp}_{L}$ , which are switched on
simultaneously.
\begin{eqnarray}
\fl\Gamma(\omega)=\frac{2\pi\sqrt{\varepsilon(\omega)}}{c\hbar
N_{\bf
f}}\left[1-\exp\left(-\frac{\hbar\omega}{T}\right)\right]\nonumber\\
\fl\times\sum_{n,m,n'm'} f_0(\varepsilon_{nm})\left|\langle
n'm'|\tilde{H}|nm\rangle\right|^2
\delta\left(\varepsilon_{n'm'}-\varepsilon_{nm}\mp\hbar\omega_q\mp
g\mu_0B-\hbar\omega\right)
\end{eqnarray}
Here
$\varepsilon_{nm}=\hbar\omega_1(n+1/2)+\hbar\omega_3(m+1/2)-m\alpha^2/\hbar^2,
$ ($n,m=0,1,2\ldots$), $\alpha$ is the strength of
delta-potential, $g$ is $g$-factor, $\mu_0=e\hbar/2m_0c$ is the
Bohr magneton, $\varepsilon(\omega)$ is the real part of the
dielectric constant, $\bf f$ is the wave vector of photons,
$f_0(nm)$ is the electron distribution function normalized to
unity, $N_{\bf f}$ is the number of initial-state photons,
$\omega_q$ is the phonon frequency, the factor
$(1-\exp(-\hbar\omega/T)$ takes into account spontaneous
transitions, and $\omega_i$ are the hybrid frequencies
\begin{eqnarray}
\omega_{1,3}=\frac{1}{2}\left[\sqrt{\left(\Omega_x+\Omega_z\right)^2+\omega_c^2}\pm
\sqrt{\left(\Omega_x-\Omega_z\right)^2+\omega_c^2}\right]
\end{eqnarray}

The matrix elements of the operator $\tilde{H}$ that is
responsible for spin-flip processes is determined by the formula
\begin{eqnarray}
\label{ver} \langle
n'm'|\tilde{H}|nm\rangle&=&\sum_{n''m''}\frac{\langle
n'm'|H_R|n''m''\rangle\langle
n''m''|H^{sp}_{L}|nm\rangle}{\varepsilon_{n'
m'}-\varepsilon_{n''m''}-\hbar\omega}\nonumber\\
\label{ver} &+&\sum_{n''m''}\frac{\langle
n'm'|H^{sp}_{L}|n''m''\rangle\langle
n''m''|H_R|nm\rangle}{\varepsilon_{n'm'}-\varepsilon_{n''
m''}+\hbar\omega}.
\end{eqnarray}
In (\ref{ver}), the first term describes processes involving,
first, emission of a phonon and then, absorption of a photon; and
the second term accounts for the processes involving, first,
absorption of a phonon and, then, emission of a photon.

To calculate the matrix elements of operators electron-phonon and
electron-photons interactions we use the method of a linear
canonical transformation of the phase space of the system
\cite{Gey05,Gal04,Mar06} because in this case we can analytically
calculate matrix elements of the corresponding operators.

The matrix elements of the electron-photon interaction operator
was calculated in \cite{Gal04} and have the form
\begin{eqnarray}
\langle
n''m''|H_R|nm\rangle&=&\frac{e\varepsilon_\omega\sqrt{\hbar}}{\sqrt{2m^*}\omega}
\left[X_1\sqrt{\frac{n''+1}{2}}\delta_{n',n''+1}\delta_{m',m''}\right.\nonumber\\
&+&\left.X_3\sqrt{\frac{m''+1}{2}}\delta_{n',n''}\delta_{m',m''+1}\right]. 
\end{eqnarray}
Here, $\varepsilon_\omega$ is the amplitude of the electromagnetic
wave polarized along the $Oz$ axis and the coefficients $X_i$ are
given by
\begin{eqnarray}
X_i=\frac{\Omega_z^2\omega_c}{\sqrt{\omega_i}\sqrt{\left(\Omega_z^2-\omega_i^2\right)^2+\Omega_z^2\omega_c^2}},\quad
i=1,3.
\end{eqnarray}

The part of operator of electron-phonon interaction  that is
responsible for spin-flip processes has the form
\cite{Mat67,Pav65}
\begin{eqnarray}
H^{sp}_L&=&\sum_{\bf
q}d\left(\frac{1}{2NM\omega_0\hbar}\right)^{1/2}
\left(%
\begin{array}{cc}
  0  & {[{\bf h}_{-} \times {\bf e}]} \\
  {[{\bf h}_{+}\times {\bf e}]}  & 0  \\
\end{array}%
\right)\nonumber\\
&\times&\left[e^{i{\bf qr}}b_{\bf q}\left({\bf p}+\frac{e}{c}{\bf
A}+\frac{\hbar {\bf q}}{2}\right)+e^{-i{\bf qr}}b^{+}_{\bf
q}\left({\bf p}+\frac{e}{c}{\bf A}-\frac{\hbar {\bf
q}}{2}\right)\right].
\end{eqnarray}
Here ${\bf h}_{\pm}={\bf l}_x\pm i{\bf l}_y$, ${\bf l}_x$, ${\bf
l}_y$ are the unit vectors of axes $Ox$ and $Oy$, ${\bf q}$ is the
polarization vector of an optical phonon, $M=M_1M_2/(M_1+M_2)$ is
the reduced mass of the lattice cell of III-V compounds, $d$ is
electron-phonon coupling constant, $\bf q$ is the phonon wave
vector, $N$ is the number of cells in the crystal, ${\bf
A}=(Bz/2,0,-Bx/2)$ is the vector potential of a magnetic field
$\bf B$.

Let us consider the transitions from the state with $s=-1$ to the
state with $s=1$ ($s=\pm1$ is the spin quantum number). Denoting
$D_{\bf q}=d/\sqrt{2NM\omega_0\hbar}$, we get that the part $H_L$
of the $H^{sp}_L$ operated on the coordinate part of wave
functions is
\begin{eqnarray}
H_L=\sum_{\bf q}D_{\bf q}(ie_{\parallel}{\bf i}-e_{\parallel}{\bf
j}-ie_{\perp}e^{-i\varphi}{\bf k})\left[e^{-i{\bf qr}}b^{+}_{\bf
q}\left({\bf p}+\frac{e}{c}{\bf A}-\frac{\hbar {\bf
q}}{2}\right)+\mathrm{c.c.}\right],
\end{eqnarray}
where the phonon wave vector is written in cylindrical coordinates
$e_z=e_{\parallel}$, $e_y=e_{\perp}\sin\varphi$,
$e_y=e_{\perp}\cos\varphi$.

After some cumbersome algebra, using the method of canonical
transformation of the phase space we get the following formulae
for the matrix elements of $H_L$
\begin{eqnarray}
\fl\langle n'',m''|H_L|n,m\rangle=\sum_{\bf q}D_{\bf
q}\sqrt{N_0+\frac{1}{2}\pm\frac{1}{2}}\left[C_1J(n'',m'',n,m)+C_2\sqrt{\frac{n}{2}}J(n'',m'',n-1,m)\right.\nonumber\\
+C_3\sqrt{\frac{n+1}{2}}J(n'',m'',n+1,m)+C_4\sqrt{\frac{m}{2}}J(n'',m'',n,m-1)\nonumber\\
\left.+C_5\sqrt{\frac{m+1}{2}}J(n'',m'',n,m+1)\right]
\end{eqnarray}
Here
\begin{eqnarray}
\fl
J(n'',m'',n,m)=\frac{n!m!}{n''!m''!}(-1)^{n''-n}(-1)^{m''-m}e^{-g^2/2}
e^{(\kappa_1\lambda_1+\kappa_3\lambda_3)i/2}e^{-i\varphi_1(n''-n)}e^{-i\varphi_3(m''-m)}\nonumber\\
\times
L^{n''-n}_{n}\left(g_1^2\right)L^{m''-m}_{m}\left(g_3^2\right)g_1^{n''-n}g_3^{m''-m},
\end{eqnarray}
$L^{n'}_n$ are generalized Laguerre polynomials, $\kappa_i$,
$\lambda_i$, $\varphi_i$, $g_i$ are the functions of the
characteristic frequencies of the parabolic potential, phonon wave
vector and magnetic field (see \cite{Gal04}). The constants $C_i$
are determined by the following expressions
\begin{eqnarray}
\fl C_1=-\frac{1}{2}i\hbar q_xe_{\parallel}+\frac{1}{2}i\hbar
q_ze^{-i\varphi}e_{\perp}-\hbar
q_y\frac{\mu}{\kappa_0}e_{\parallel},
\end{eqnarray}
\begin{eqnarray}
\fl C_{2}=i\mu a_{13}e_{\parallel}-\hbar \mu
a_{21}l_1e^{-i\varphi}e_{\perp}+\frac{1}{2}im^{*}\omega_c\mu
a_{43} e_{\parallel}+\frac{1}{2}\hbar m^{*}\omega_c\mu
a_{31}l_1e^{-i\varphi}e_{\perp},
\end{eqnarray}
\begin{eqnarray}
\fl C_{3}=i\mu a_{13}e_{\parallel}+\hbar \mu
a_{21}l_1e^{-i\varphi}e_{\perp}+\frac{1}{2}im^{*}\omega_c\mu
a_{43} e_{\parallel}-\frac{1}{2}\hbar m^{*}\omega_c\mu
a_{31}l_1e^{-i\varphi}e_{\perp},
\end{eqnarray}
\begin{eqnarray}
\fl C_{4}=i\mu a_{14}e_{\parallel}-i\hbar \mu
a_{22}l_1e^{-i\varphi}e_{\perp}+\frac{1}{2}im^{*}\omega_c\mu
a_{44} e_{\parallel}+\frac{1}{2}\hbar m^{*}\omega_c\mu
a_{32}l_1e^{-i\varphi}e_{\perp},
\end{eqnarray}
\begin{eqnarray}
\fl C_{5}=i\mu a_{14}e_{\parallel}-i\hbar \mu
a_{22}l_1e^{-i\varphi}e_{\perp}+\frac{1}{2}im^{*}\omega_c\mu
a_{44} e_{\parallel}-\frac{1}{2}\hbar m^{*}\omega_c\mu
a_{32}l_1e^{-i\varphi}e_{\perp}.
\end{eqnarray}
Here $a_{ji}$ are the matrix elements of the transition matrix
from the old phase variables $(p_x,p_y,p_z,x,y,z)$ to the new
phase variables $(P_1,P_2,P_3,Q_1,Q_2,Q_3)$ \cite{Gal04},
$\kappa_0=m^*\alpha/\hbar^2$.

Let us consider in (\ref{ver}) processes with emission of a phonon
and absorption of a photon. In this case we can write the
absorption coefficient as a sum of partial absorptions
\begin{eqnarray}
\Gamma(\omega)=\sum_{n'm',n'm}\Gamma(n',m',n',m),
\end{eqnarray}
where
\begin{eqnarray}
\label{fin}
\fl&&\Gamma(n',m',n',m)=\frac{e\varepsilon_\omega\sqrt{\hbar}\sqrt{\varepsilon}}{2\sqrt{m^*}cN_{\bf
f}\omega}\left[1-\exp\left(-\frac{\hbar\omega}{T}\right)\right]f_0(\varepsilon_{nm})\nonumber\\
&\times&\sum_{\bf
q}\left|D_q\sqrt{N_q+1}\left\{\frac{X_1}{\omega_1-\omega}\left[\sqrt{n'}C_1J(n'-1,m',n,m)+\sqrt{n}C_2J(n'-1,m',n-1,m)
\right.\right.\right.\nonumber\\
&+&\sqrt{n+1}C_3J(n'-1,m',n+1,m)+\sqrt{m}C_4J(n'-1,m',n,m-1)\nonumber\\
&+&\sqrt{m+1}C_5J(n'-1,m',n,m+1)+\frac{X_3}{\omega_3-\omega}\left[
\sqrt{n'}C_1J(n',m'-1,n,m)\right.\nonumber\\
&+&\sqrt{n}C_2J(n',m'-1,n-1,m)+\sqrt{n+1}C_3J(n',m'-1,n+1,m)\nonumber\\
&+&\left.\left.\left.\sqrt{m}C_4J(n',m'-1,n,m-1)+\sqrt{m+1}C_5J(n',m'-1,n,m+1)\right]\right\}\right|^2
\delta\left(\Delta\omega\right),\nonumber\\
\end{eqnarray}
where $\Delta\omega=\omega_1(n'-n)+\omega_3(m'-m)+\omega\pm g\mu_0
B-\omega_q$ is the resonance detuning.

In the case of transitions which originate from the states $(0,0)$
into $(0,0)$ because these transitions give the main contribution
in the absorption. In this case we get
\begin{eqnarray}
\label{fin}
\fl&&\Gamma(0,0,0,0)=\frac{e\varepsilon_\omega\sqrt{\hbar}\sqrt{\varepsilon}}{2\sqrt{m^*}cN_{\bf
f}\omega}\left[1-\exp\left(-\frac{\hbar\omega}{T}\right)\right]f_0(\varepsilon_{00})\nonumber\\
&\times&\sum_{\bf
q}\left|D_q\sqrt{N_q+1}\left(C_3+C_5\right)\left\{\frac{X_1}{\omega_1-\omega}
 +\frac{X_3}{\omega_3-\omega}\right\}\right|^2
\delta\left(\Delta\omega\right),\nonumber\\
\end{eqnarray}

 Replacing the sum by the integral and assuming a
parabolic dispersion law for long-wave phonons
$\omega_q=\omega_0(1-\omega_0^{-2}V_sq^2)$, where $\omega_0$ is
the optical-phonon threshold frequency and $V_s$ is the speed of
sound, one can easily evaluate the integral with respect to $|{\bf
q}|$ (converting to spherical coordinates) thanks to the presence
of a delta function.

In conclusion, we have investigated theoretically the
spin-hybrid-phonon resonance in anisotropic quantum dots in the
presence of a magnetic field. If we ignore optical phonon
dispersion, the partial absorption peaks have a delta-function
singularity at the points where $\Delta\omega=0$.  Hence, in this
case arise additional resonances in the small vicinity of the
peaks of the hybrid-phonon resonances \cite{Mar02} due to
spin-flip processes. These peaks are symmetrically positioned to
the left and right to the points of hybrid-phonon resonances. The
width and the position of the resonance peaks depend strongly on
the magnetic field and the characteristic frequencies of the
parabolic confinement.

\ack The present work was supported by the Russian Foundation for
Basic Research, project no. 05-02-16145.

\end{document}